\documentclass[letterpaper,11pt]{article}

\usepackage{graphicx}
\usepackage{amsmath}

\begin{document}

\title{\textbf{Equalization of Response Functions of the SK and SNO}}
\author{Faisal Akram and Haris Rashid \\
{\small Centre for High Energy Physics}\\
{\small University of the Punjab}\\
{\small Lahore, Pakistan}} \maketitle

\begin{abstract}

We have studied the equalization of the response functions of the
SK and SNO for total event rate and event rate for the energy
bins. To calculate the response functions of SNO, we have used the
latest theoretical values of the cross section of the
neutrino-deuteron CC process. By using these new theoretical
values, we find that the trigger threshold of the SK at which its
response function is equalized to the response function of SNO (at
the trigger threshold of 6.75 MeV), is 8.5 MeV. This value is
0.1~MeV smaller than the value calculated using the old
theoretical values of the cross section of neutrino-deuteron
process. The use of these new theoretical values also produces a
small change in the range of the energy bins where the response
functions are equalized.

\end{abstract}

\bigskip
\bigskip

\noindent Faisal Akram

\noindent Center for High Energy Physics

\noindent University of the Punjab

\noindent Lahore, Pakistan

\noindent Phone: 92-42-9231137

\noindent email: faisal@chep.pu.edu.pk

\pagebreak

\section{Introduction}

Homestake, GALLEX, SAGE, Kamiokande, Super-Kamiokande and SNO
[1-6], these are the six solar neutrino experiments for which up
to date experimental results are available. Out of these six
detectors, first three are radiochemical detectors and can measure
only the total event rate produced by the electron neutrinos,
where as other three are water-cherenkov detectors and provide the
facility to record the time of the event, energy of the scattered
electron and scattering angle. These detectors are sensitive to
all flavors of neutrinos. All these detectors have detected solar
neutrinos at the rate much smaller than predicted by the Standard
Solar Model (SSM) [7]. This discrepancy between SSM and
experimental measurements is called solar neutrino puzzle (SNP).
The puzzle can be explained in two general ways.

1. Perhaps the solar models do not accurately describe the sun.

2. Perhaps the known theories of neutrino do not accurately
describe it.

At present it is difficult to accept (1) because the SSM has been
very successful in describing many features of the sun,
particularly the latest confirmation of the SSM's prediction of
the velocity of sound at the surface of sun through
helioseismological studies [7]. Now it is generally believed that
the solution of the puzzle is to come from the particle theory.
The most elegant solution, which the particle physics provides, is
the 'neutrino oscillation'. The phenomena in which different
flavors of neutrino may oscillate into each other while passing
through vacuum or matter [8]. The exact amount of depletion which
may be caused by the oscillations, however, depends on the
neutrinos squared mass differences and mixing angles. The 'Global
analysis of solar neutrino data', in which we fix the values of
these parameters so that the difference between the predicted and
measured rates is minimum, reveals that there are many possible
solutions in the frame work of neutrino oscillations [9]. However,
these solutions have different goodness-of-fit (g.o.f). The latest
analysis shows that the Large Mixing Angle (LMA) solution have the
highest g.o.f and hence most probable [9].

There is another mode of analysis in which we try to probe the
oscillations from the experimental data, without using the
prediction of solar neutrinos flux by the SSM. In this model
independent analysis we compare the data in the same or different
solar neutrinos experiments and try to find out whether it is
consistent with the neutrino oscillations or not [10]. This
comparison of the solar neutrino data obtained from different
detectors, however, requires the equalization of their response
functions. The equalization of the response functions of two
different detectors is non-trivial phenomena and so far it has
been possible only for the SK and SNO detectors. In [10,11,12] the
authors have proved that it is possible to equalize the response
functions of the SK and SNO by changing the trigger threshold or
energy bins of one of these detectors. The comparison, after the
equalization of the response functions, reveals that the results
obtained from the SK and SNO are consistent with the neutrino
oscillations[6,10].

In the equalization of these response functions, accurate values
of the cross sections of neutrino-electron ($ve$) and
neutrino-deuteron ($vd$) reactions, occurring in the SK and SNO
respectively, are very important. The most accurate calculations
of the cross sections of $ve$ process is given in [16]. These
calculations include the effect QED and QCD radiative corrections
up to one-loop. For neutrino-deuteron reactions, the most
successful method used is the standard nuclear physics approach
(SNPA) based on one-body impulse approximation terms and two-body
exchange-current terms acting on non-relativistic nuclear wave
function. The status of the first detailed calculation based on
SNPA is given in [17], referred as Kubodera-Nozawa~(KN)
calculations. The calculated values of the cross section of $vd$
process in [17] have been used in the equalization of response
functions of the SK and SNO in [10,11,12]. The work of Kubodera
and Nozawa was further improved by using more accurate NN
potentials and nucleon weak-interaction form factors [18]. In the
literature this improved work is referred as the NSGK
calculations. The estimates of NSGK calculations have been further
improved by about 1\% by updating some of its inputs and by taking
into account the results of recent effective-field-theoretical
calculation [19].

In our work we have studied the response functions of the SK and SNO detectors for B$%
_{8}$ neutrinos using the most accurate estimate of cross sections
of $ve$ and $vd$ process taken from reference [16] and [19]
respectively. In section 2 we present the basic definitions of the
response functions. In section 3 and 4 we present our results of
the comparison of the response functions. As a consequence of the
approximate equalization of the response functions, we show in
section 5, that in the presence of active neutrino oscillations,
how the total event rates and event rates for energy bins of the
SK and SNO detectors are related to each other.

\section{SK and SNO Response Functions}

The solar neutrinos are detected in the SK by the following
elastic scattering (ES) process.
\begin{equation}
v_{e,a}+e^{-}\longrightarrow v_{e,a}+e^{-}
\end{equation}

This process is caused either by the electron neutrino $(v_{e})$
or other two
active neutrinos $v_{a}(a=\mu ,\tau )$ with cross sections $\sigma ^{e}$and $%
\sigma ^{a}$ respectively.

In the SNO, the neutrinos are detected by the following
charged-current (CC) and neutral-current (NC) processes.
\begin{eqnarray}
v_{e}+d &\longrightarrow &p+p+e^{-}\text{ \ \ \ \ \ \ \ \ \ \ \ \ (CC
Process)} \\
v_{e, \mu, \tau}+d &\longrightarrow &p+n+e^{-}\text{ \ \ \ \ \ \ \
\ \ \ \ \ (NC Process)}
\end{eqnarray}

The CC process is caused by only electron neutrinos with cross section $%
\sigma ^{c},$where as NC process is caused by the all three
flavors of neutrinos with the same cross section $\sigma ^{n}.$

All the scattering events in the SK and SNO are detected by
detecting the cherenkov radiation emitted by the scattered
electron along its direction of motion. This method allows us to
record the energy as well as the direction of the scattered
electrons. However, due to finite detector resolution the measured
energy $(E_e)$ is expected to be scattered around its actual value
$(E'_e)$ according to some well-defined distribution function,
called resolution of the detector [13-15]. In this way it is
possible to define the total event rates by applying some limit on
the minimum value of the measured energy, called trigger threshold
and event rate for a measured energy range, called energy bin.
Normally the trigger threshold and energy bin are taken not less
than 5-6 MeV to reduce the background effects. At this higher
value of threshold only B$_{8}$\ or possibly hep neutrinos
contribute to the measured events.

The response function for the total event rate (or the rate for an
energy bin) is the normalized effective cross section for
producing the scattered electron with the measured energy greater
than the trigger threshold E$_{th}$(or lying in a specific energy
bin) [11,12,14]. The response functions (RF) relevant for our work
are
\begin{description}
\item $\rho _{E_{th}}^{e}(E_{v})$, the RF of ES $(v_{e},e)$ for event
rate above $E_{th}$

\item $\rho _{E_{th}}^{a}(E_{v})$, the RF of ES $(v_{a},e)$ for event rate above $%
E_{th}$

\item $\rho _{E_{th}}^{c}(E_{v})$, the RF of CC $(v_{e},e)$ for event rate above $%
E_{th}$

\item $\rho _{i}^{e}(E_{v})$, the RF of ES $(v_{e},e)$ for event rate
in $ith$ energy bin

\item $\rho _{i}^{a}(E_{v}$, the RF of ES $(v_{a},e)$ for event rate
in $ith$ energy bin

\item $\rho _{i}^{c}(E_{v})$, the RF of CC $(v_{e},e)$ for event rate
in $ith$ energy bin
\end{description}
\noindent These response functions are defined as following.
\begin{eqnarray}
\rho _{E_{th}}^{e}(E_{v}) &=&\frac{\lambda _{B}(E_{v})\int_{E_{th}}^{E_{\max
}}dE_{e}\int_{0}^{E_{v}}dE_{e}^{\prime }\frac{d\sigma
^{e}(E_{v},E_{e}^{\prime })}{dE_{e}^{\prime }}R_{SK}(E_{e},E_{e}^{\prime })}{%
\sigma _{E_{th}}^{e}} \\
\rho _{E_{th}}^{a}(E_{v}) &=&\frac{\lambda _{B}(E_{v})\int_{E_{th}}^{E_{\max
}}dE_{e}\int_{0}^{E_{v}}dE_{e}^{\prime }\frac{d\sigma
^{a}(E_{v},E_{e}^{\prime })}{dE_{e}^{\prime }}R_{SK}(E_{e},E_{e}^{\prime })}{%
\sigma _{E_{th}}^{a}} \\
\rho _{E_{th}}^{c}(E_{v}) &=&\frac{\lambda _{B}(E_{v})\int_{E_{th}}^{E_{\max
}}dE_{e}\int_{0}^{E_{v}}dE_{e}^{\prime }\frac{d\sigma
^{c}(E_{v},E_{e}^{\prime })}{dE_{e}^{\prime }}R_{SK}(E_{e},E_{e}^{\prime })}{%
\sigma _{E_{th}}^{c}} \\
\rho _{i}^{e}(E_{v}) &=&\frac{\lambda _{B}(E_{v})\int_{E_{\min
}^{i}}^{E_{\max }^{i}}dE_{e}\int_{0}^{E_{v}}dE_{e}^{\prime }\frac{d\sigma
^{e}(E_{v},E_{e}^{\prime })}{dE_{e}^{\prime }}R_{SNO}(E_{e},E_{e}^{\prime })%
}{\sigma _{i}^{e}} \\
\rho _{i}^{a}(E_{v}) &=&\frac{\lambda _{B}(E_{v})\int_{E_{\min
}^{i}}^{E_{\max }^{i}}dE_{e}\int_{0}^{E_{v}}dE_{e}^{\prime }\frac{d\sigma
^{a}(E_{v},E_{e}^{\prime })}{dE_{e}^{\prime }}R_{SK}(E_{e},E_{e}^{\prime })}{%
\sigma _{i}^{a}} \\
\rho _{i}^{c}(E_{v}) &=&\frac{\lambda _{B}(E_{v})\int_{E_{\min
}^{i}}^{E_{\max }^{i}}dE_{e}\int_{0}^{E_{v}}dE_{e}^{\prime }\frac{d\sigma
^{c}(E_{v},E_{e}^{\prime })}{dE_{e}^{\prime }}R_{SNO}(E_{e},E_{e}^{\prime })%
}{\sigma _{i}^{e}}
\end{eqnarray}

Where the differential cross sections $\frac{d\sigma
^{e}}{dE_{e}^{\prime }}$ and $\frac{d\sigma ^{a}}{dE_{e}^{\prime
}}$ for ES processes are taken from ref [16] and $\frac{d\sigma
^{c}}{dE_{e}^{\prime }}$ for CC process is taken from the ref
[19]. It is noted that in [10,11,12] the authors have used old
theoretical data for CC process [17]. $R_{SK}$ and $R_{SNO}$ are
the resolutions functions of the SK and SNO respectively. $\lambda
_{B}(E_{v})$ is the normalized energy spectrum of B$^{8}$
neutrinos. The denominators represent the B$^{8}$ neutrinos total
cross sections for producing an electron with the observed energy
greater than the trigger threshold or lying with in the specific
energy bin as per the definition of response function. The
denominator in each expression is obtained by integrating the
numerator over the energy $(E_v)$ of B$^8$ neutrinos.

\section{Equalization of the Total Response Functions}

In this section, it is shown that the following total response
functions for the SK and SNO can be equalized to a good
approximation by an appropriate choice of the trigger threshold.
\begin{eqnarray}
\rho _{E_{th}}^{e}(E_{v}) &=&\rho _{E_{th}}^{a}(E_{v})  \label{e1} \\
\rho _{E_{th}^{\prime }}^{e}(E_{v}) &\simeq &\rho _{E_{th}}^{c}(E_{v})
\label{e2}
\end{eqnarray}
\qquad

Equalization of ES response functions of the SK for $v_{e}$ and
$v_{a}$ neutrinos is accurately possible where as the ES response
function of the SK for
$v_{e}$ can be approximately equalize to the CC response function of SNO for $%
v_{e}$ by changing the trigger threshold of either of the detectors. In
order to compare the response functions, we determine the following integral
differences.
\begin{eqnarray}
\Delta _{1} &=&\int dE_{v}|\rho _{E_{th}}^{e}(E_{v})-\rho
_{E_{th}}^{a}(E_{v})| \\
\Delta _{2} &=&\int dE_{v}|\rho _{E_{th}^{\prime }}^{e}(E_{v})-\rho
_{E_{th}}^{c}(E_{v})|
\end{eqnarray}

\begin{figure}[p]
\centering
\includegraphics[width=0.75\textwidth ]{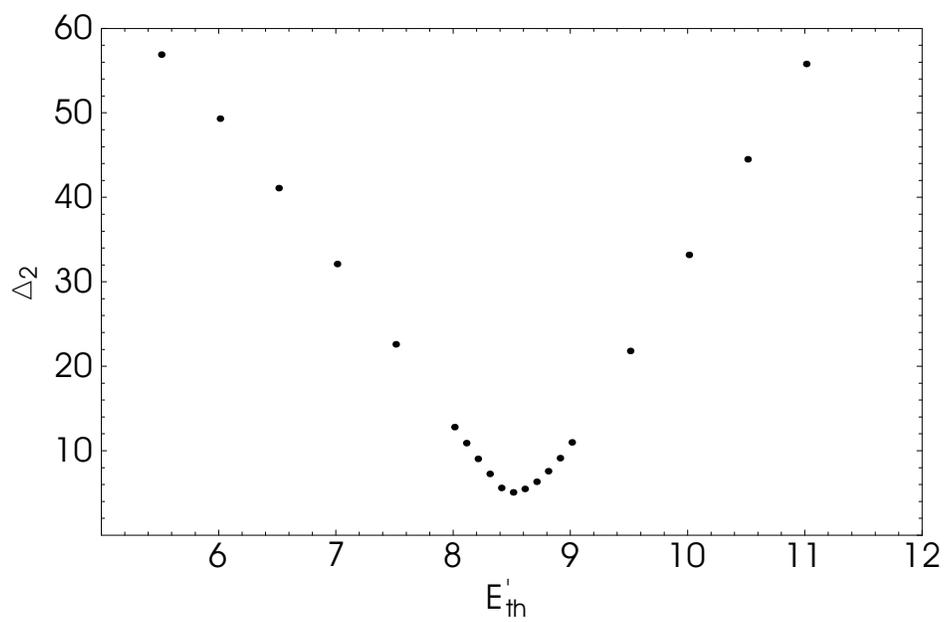}
\caption{The value of square integral difference of total response
functions at different value of trigger threshold of SK}
\label{g1}
\end{figure}

\begin{figure}[p]
\centering
\includegraphics[width=0.75\textwidth ]{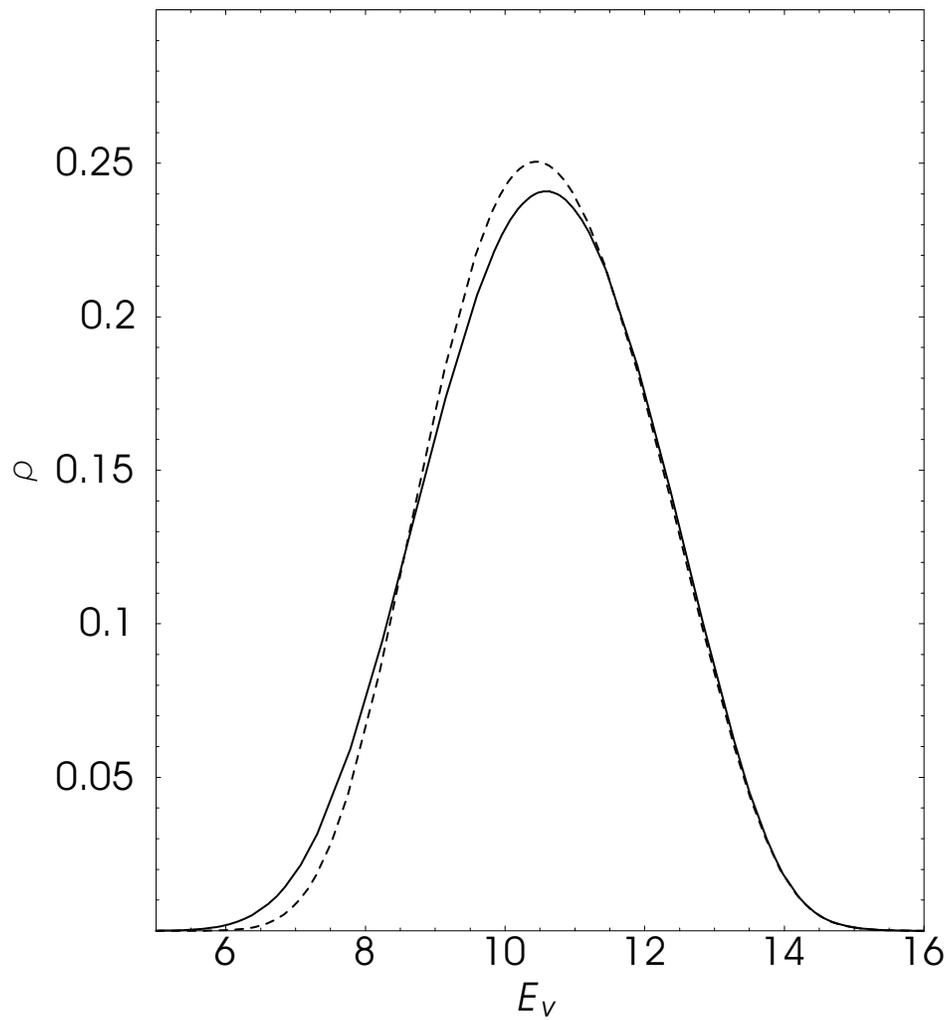}
\caption{Best equalization of total response functions of SK
(continuous line) and SNO (dotted line)} \label{g2}
\end{figure}

By requiring the minimization of these integral differences we
obtain the best equalization of the response functions. These
integral difference are zero if the equations \ref{e1} and
\ref{e2} satisfy exactly. It is found that the first integral
difference is zero at all values of $E_{th}$. In
minimizing the second integral difference, we take SNO's threshold, $%
E_{th}=6.75$ MeV and obtain the value of $\Delta _{2}$ for different values
of SK's threshold $E_{th}^{\prime }$. The results in Figure \ref{g1} shows that $%
\Delta _{2}$ has one minima at $E_{th}^{\prime }=8.5$ MeV. At this
value of SK's threshold, the response function are best equalized.
Figure \ref{g2} shows the graphs of these approximately equalized
response functions of the SK and SNO. The value of $E_{th}^{\prime
}$ obtained by the old theoretical data of CC cross section in
[10,11], is 8.6 MeV. In the combined SK and SNO data analysis by
SNO collaboration in [6], the value of 8.5 MeV was used, which is
in agreement with our calculation.

\section{Equalization of the Response Functions for Energy Bins}

In this section, it is shown that the following response functions
of the SK and SNO for the energy bins can be equalized to a good
approximation by an appropriate choice of the range of energy
bins.
\begin{eqnarray}
\rho _{i}^{e}(E_{v}) &=&\rho _{i}^{a}(E_{v}) \\
\rho _{i}^{e}(E_{v}) &\simeq &\rho _{i^{\prime }}^{c}(E_{v})
\end{eqnarray}
\qquad

Equalization of the ES response functions of SK for $v_{e}$ and
$v_{a}$ neutrino is accurately possible for the same energy bins
where as the CC response function of SNO for $v_{e}$ can be
approximately equalized to ES response function of SK for $v_{e}$
for different energy bins. In order to compare these response
functions, we determine the following integral differences.
\begin{eqnarray}
\Delta _{1} &=&\int dE_{v}|\rho _{i}^{e}(E_{v})-\rho _{i}^{a}(E_{v})| \\
\Delta _{2} &=&\int dE_{v}|\rho _{i}^{e}(E_{v})-\rho _{i^{\prime
}}^{c}(E_{v})|
\end{eqnarray}

\begin{figure}[p]
\centering
\includegraphics[width=0.75\textwidth ]{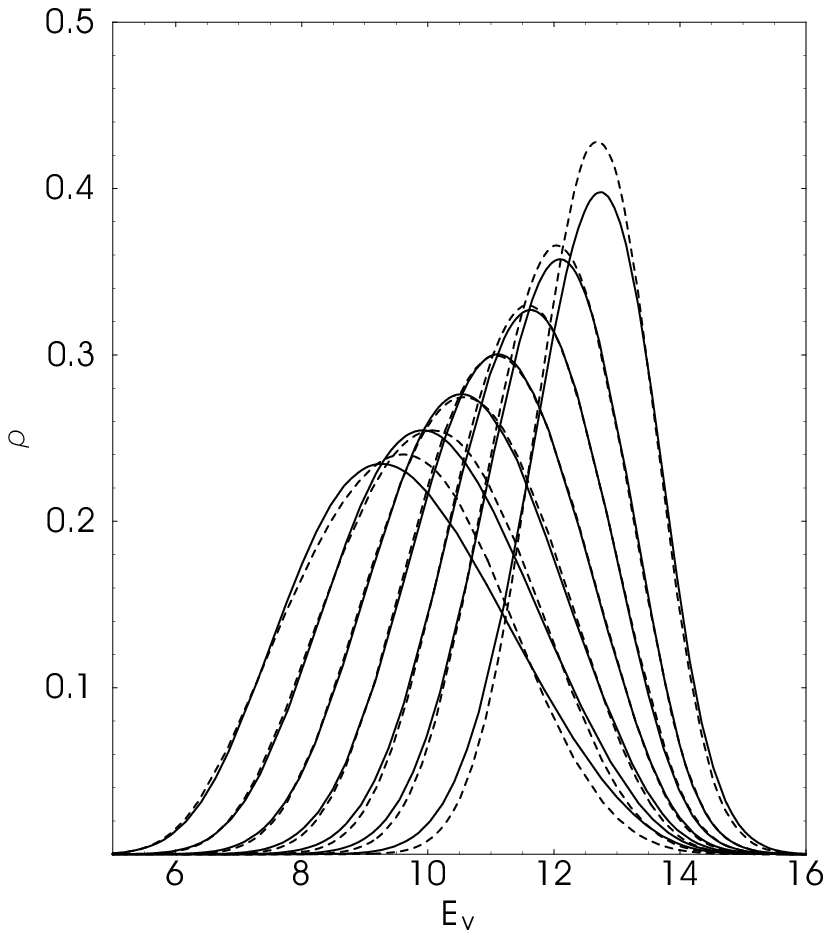}
\caption{Equalized response functions of SK (continuous curve) and
SNO (dotted curve) for odd energy bins}
\label{g3}
\end{figure}

\begin{figure}[p]
\centering
\includegraphics[width=0.75\textwidth ]{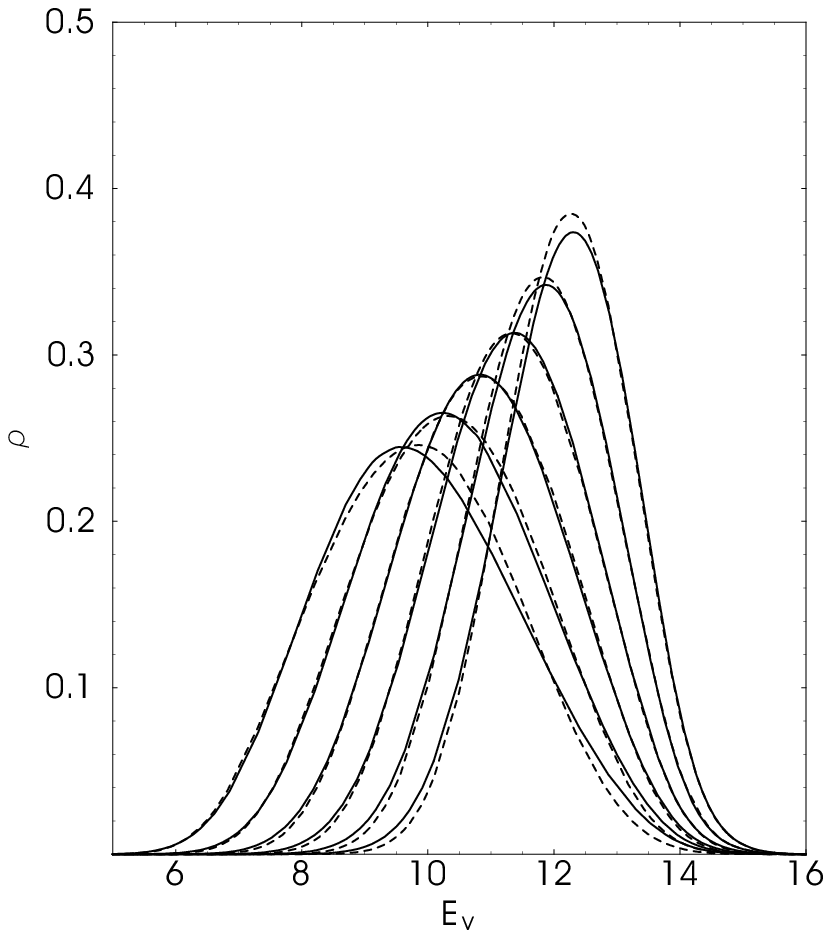}
\caption{Equalized response functions of SK (continuous curve) and
SNO (dotted curve) for even energy bins}
\label{g4}
\end{figure}

By requiring the minimization of these integral differences, we
obtain the best equalization of the response functions for energy
bins. It is found that the first integral difference is zero for
all energy bins of SK. In minimizing the second integral
difference, we take 13 different energy bins
of the SK for which the experimental data is available and obtain the values of $%
\Delta _{2}$ for different energy bins of SNO. For each energy bin
of SK, we find the energy bin of SNO for which $\Delta _{2}$ is
minimum. The results are summarized in the Table \ref{t1}. Figure
\ref{g3} and \ref{g4} shows the graphs of these approximately
equalized response functions for energy bins of the SK and SNO.

\section{Relation between SK and SNO Event Rates}

After equalizing the total and energy bin response functions of
the SK and SNO, it is possible to relate the corresponding event
rates [10,11,12]. First we relate the total event rate defined
above the trigger threshold. These total event rate, assuming no
oscillation, are given by.
\begin{eqnarray}
R_{SK}^{0} &=&\phi \sigma _{E_{th}}^{e} \\
R_{SNO}^{0} &=&\phi \sigma _{E_{th}}^{c}
\end{eqnarray}

where $\phi =5.15\times 10^{6}cm^{-2}s^{-1},$ is the SSM's
predicted total flux of B$^{8}$ neutrinos and $\sigma
_{E_{th}}^{e},$ $\sigma _{E_{th}}^{c}$ are total effective cross
section for observing the event above the trigger threshold of SK
and SNO respectively. In the presence of neutrino
oscillations, which are described by a survival probability function $%
P_{ee}(E_{v}),$ the expressions for the total event rates of the
SK and SNO are given by
\begin{eqnarray}
R_{SK} &=&\phi \lbrack \sigma _{E_{th}}^{e}<P_{ee}>_{E_{th}}^{e}+\sigma
_{E_{th}}^{a}(1-<P_{ee}>_{E_{th}}^{a})] \\
R_{SNO} &=&\phi \lbrack \sigma _{E_{th}}^{c}<P_{ee}>_{E_{th}}^{c}
\end{eqnarray}

where the terms $<P_{ee}>_{E_{th}}^{x}(x=e,a,c)$ represent the
average survival probability weighted by the response functions
$\rho _{E_{th}}^{e}$, $\rho _{E_{th}}^{a}$ and $\rho
_{E_{th}}^{c}$. Now if these response functions are equal, as they
are by the equations 14 and 15, then it implies the equalization
of these three response function's weighted survival
probabilities.
\begin{equation}
<P_{ee}>_{E_{th}}^{e}=<P_{ee}>_{E_{th}}^{a}\approx
<P_{ee}>_{E_{th}}^{c}\equiv <P_{ee}>
\end{equation}

The event rates normalized to un-oscillated rates are now given by
\begin{eqnarray}
r_{SK} &\equiv &\frac{R_{SK}}{R_{SK}^{0}}=[<P_{ee}>+\frac{\sigma
_{E_{th}}^{a}}{\sigma _{E_{th}}^{e}}(1-<P_{ee}>)] \\
r_{SNO} &\equiv &\frac{R_{SNO}}{R_{SNO}^{0}}=<P_{ee}>
\end{eqnarray}

Eliminating $<P_{ee}>$ in equations 23 and 24, we get the
following relation between normalized event rates of the SK and
SNO.
\begin{equation}
r_{SK}=r_{SNO}(1-\frac{\sigma _{E_{th}}^{a}}{\sigma _{E_{th}}^{e}})+\frac{%
\sigma _{E_{th}}^{a}}{\sigma _{E_{th}}^{e}}
\end{equation}

The value of the ratio $\frac{\sigma _{E_{th}}^{a}}{\sigma _{E_{th}}^{e}}$
is calculated to be 0.1518.

Similarly by applying the equalization of the response functions
of the SK and SNO for energy bins, we can obtain the similar
relation between the SK and SNO event rates for energy bins.
\begin{equation}
r_{SK}^{i}=r_{SNO}^{i}(1-\frac{\sigma _{i}^{a}}{\sigma _{i}^{e}})+\frac{%
\sigma _{i}^{a}}{\sigma _{i}^{e}}
\end{equation}

where $r_{SK}^{i}$ and $r_{SNO}^{i}$ are the normalized event
rates of $ith$ energy bin of SK and SNO respectively and $\sigma
_{i}^{e,a}$ are the effective cross sections for producing the
event in the $ith$ energy bin by $v_{e}$ and $v_{a}$ neutrinos.
The values of these cross sections are given in Table\ 1.

\section{Conclusion}
In this work, we have shown that the SK and SNO response functions
of ES and CC processes can be approximately equalized with the new
theoretical values of the cross sections of neutrino-deuteron CC
process. Trigger threshold of the SK at which its response
function is best equalized with the response function of SNO is
found to be 8.5 MeV. This value is 0.1 MeV smaller than the values
obtained with the old values of the cross sections of
neutrino-deuteron CC process. The new values of the CC cross
sections also produce a small change in the energy bins of the SNO
for which the response functions for the energy bins are best
equalized.

\begin{table}[p]
\centering
\begin{tabular}{|lllllll|}
\hline\hline $i$ & Energy Bin & Energy Bin & $\Delta \times 100$ &
$\sigma ^{e}$ & $\sigma ^{a}$
& $\sigma ^{c}$ \\
& SK, MeV & SNO, MeV &  & (10$^{-46}$ cm$^{2})$ & (10$^{-46}$
cm$^{2})$ & (10$^{-42}$ cm$^{2})$ \\ \hline
1 & \lbrack 8.0, 8.5] & [5.10, 9.90] & 6.85 & 14.158 & 2.187 & 0.805 \\
2 & \lbrack 8.5, 9.0] & [5.60, 10.35] & 5.12 & 11.827 & 1.815 & 0.770 \\
3 & \lbrack 9.0, 9.5] & [6.10, 10.75] & 3.74 & 9.705 & 1.482 & 0.713 \\
4 & \lbrack 9.5, 10.0] & [6.60, 11.25] & 2.27 & 7.812 & 1.187 & 0.650 \\
5 & \lbrack 10.0, 10.5] & [7.10, 11.70] & 1.93 & 6.161 & 0.932 & 0.573 \\
6 & \lbrack 10.5, 11.0] & [7.60, 12.20] & 1.58 & 4.749 & 0.716 & 0.491 \\
7 & \lbrack 11.0, 11.5] & [8.05, 12.70] & 1.42 & 3.574 & 0.537 & 0.417 \\
8 & \lbrack 11.5, 12.0] & [8.50, 13.30] & 1.46 & 2.619 & 0.393 & 0.345 \\
9 & \lbrack 12.0, 12.5] & [9.00, 14.20] & 1.96 & 1.866 & 0.279 & 0.269 \\
10 & \lbrack 12.5, 13.0] & [[9.45, 15.00] & 2.40 & 1.290 & 0.193 & 0.207 \\
11 & \lbrack 13.0, 13.5] & [9.90, 18.10] & 2.95 & 0.863 & 0.129 & 0.155 \\
12 & \lbrack 13.5, 14.0] & [10.35, 18.85] & 3.45 & 0.558 & 0.083 & 0.111 \\
13 & \lbrack 14.0, 20] & [11.30, 19.50] & 7.75 & 0.812 & 0.120 & 0.483 \\
\hline\hline
\end{tabular}
\caption{Energy bins of SK (1st column), energy bins of SNO (2nd
column) where the response functions are equalizee, minumum
integral difference of response functions (3rd column) and
calculated cross sections of ES and CC processes (rest of the
columns)} \label{t1}
\end{table}

\begin{thebibliography}{99}
\bibitem{}  K. Lande, T. Daily, R. Davis, J. R. Distel, B. T. Cleveland, C.
K. Lee, P. S. Wildenhain, and J. Ullman, Astrophys. J. 496, 505 (1998); K.
Lande, P. Wildenhain, R. Corey, M. Foygel, and J. Distel, in Neutrino 2000,
p. 50

\bibitem{}  V. Gavrin for the SAGE Collaboration in Neutrino 2000, p. 36;

\bibitem{}  E. Bellotti for the GALLEX and GNO Collaborations, Neutrino
2000, p.44.

\bibitem{}  Y. Fukuda et. al., Kamiokande Collaboration, Phys. Rev. Lett.
77, 1683 (1996)

\bibitem{}  S. Fukuda et. al., Super-Kamiokande Collaboration, Phys. Rev.
Lett. 86, 5651 (2001)

\bibitem{}  SNO Collaboration, Phys. Rev. Lett. 87, 071301 (2001); SNO Collaboration, Phys. Rev. Lett. 89, 011301 (2002)

\bibitem{}  J. N. Bahcall, M. H. Pinsonneault, and S. Basu, Astrophysical
Journal, 555, 990 (2001); astro-ph/0010346 v2

\bibitem{}  B. Pontercorvo, JETP, 26, 984 (1968); L. Wolfenstein, Phys. Rev.
D17, 2369 (1978); S. P. Mikheyev and A. Yu. Smirnov, Sov. J. Nucl. Phys. 42,
913 (1985)

\bibitem{}  J. N. Bahcall, M. C. Gonzalez-Garcia, and C. Pe\~{n}a-Garay,
JHEP 207, 054 (2002)

\bibitem{}  G. L. Fogli, E. Lisi, D. Montanino, and A. Palazzo, Phys. Rev. D 64, 093007 (2001)

\bibitem{}  F. L. Villante, G. Fiorentini, and E. Lisi, Phys. Rev. D 59,
013006 (1999)

\bibitem{}  G. L. Fogli, E. Lisi, A. Palazzo, and F. L. Villante, Phys. Rev.
D 63, 113016 (2001)

\bibitem{}  J. N. Bahcall and E. Lisi, Phys. Rev. D 54, 5417 (1996)

\bibitem{}  B. Faid. G. L. Fogli, E. Lisi, and D. Montanino, Phys. Rev. B
55, 1353 (1997)

\bibitem{}  G. L. Gogli, E. Lisi, and D. Montanino, Astropart. Phys. 9, 119
(1998)

\bibitem{}  J. N. Bahcall, M. Kamionkowsky \& A. Sirlin, Phys. Rev. D 51,
6146 (1995)

\bibitem{}  K. Kubodera and S. Nozawa, Int. J. Mod. Phys. E 3, 101
(1994)

\bibitem{}  S. Nakamura, T. Sato, V. Gudkov \& K. Kubodera, Phys. Rev. C 63,
034617 (2001)


\bibitem{}  S. Nakamura, T. Sato, S. Ando, T. S. Park, F. Myhrer, V. Gudkov
and K. Kubodera, Nucl. Phys. A 707, 561 (2002)

\end{thebibliography}
\end{document}